\documentclass[aps,prl,reprint,groupedaddress,showpacs,amsmath,amssymb,twocolumn,floatfix]{revtex4-1}
\usepackage{graphicx}
\usepackage{bm}
\usepackage{color}
\newcommand{\el}{\bm{\hat{l}}}

\begin{document}

\title{Stability of superfluid $^3$He-B in compressed aerogel}

\author{J.I.A. Li}
\email[]{jiali2015@u.northwestern.edu}
\author{A.M. Zimmerman}
\author{J. Pollanen}
\thanks{Present Address: Condensed Matter Physics, California Institute of Technology, Pasadena, California 91125, USA}
\author{C.A. Collett}
\author{W.J. Gannon}
\thanks{Present Address: Department of Physics and Astronomy, Stony Brook University, Stony Brook, New York 11794, USA}
\author{W.P. Halperin}
\email[]{w-halperin@northwestern.edu}
\affiliation{Northwestern University, Evanston, IL 60208, USA}

\date{\today}

\begin{abstract}
In recent work it was shown that new anisotropic $p$-wave states of superfluid $^3$He can be stabilized within high porosity silica aerogel under uniform positive strain~\cite{Pol.12a}.  In contrast, the equilibrium phase  in an unstrained aerogel, is the isotropic superfluid B-phase~\cite{Pol.11}.  Here we report that this phase stability depends on the sign of the strain.  For negative strain of $\sim$20\% achieved by compression, the B-phase can be made more stable than the anisotropic A-phase resulting in a tricritical point for A, B, and normal phases  with a critical field of $\sim 100$ mT.  From pulsed NMR measurements we identify these phases and the orientation of the angular momentum.

\end{abstract}

\maketitle

The natural state of superfluid $^3$He is  completely free of impurities.  In zero magnetic field it has two phases, one of which has an isotropic order parameter amplitude, the B-phase, and the other, the A-phase, is axially anisotropic.  Near the superfluid transition temperature $T_c$, energetics favor the A-phase at pressures greater than 21 bar.  However, in the presence of quenched disorder from high-porosity, silica aerogel, theory indicates~\cite{Thu.98} that isotropic  scattering of the $^3$He quasiparticles  will tip the stability balance in favor of the isotropic phase.  Furthermore, it is predicted that anisotropic scattering will have the opposite effect, stabilizing anisotropic phases like the A-phase.  These predictions have been confirmed experimentally for isotropic~\cite{Ger.02, Pol.11, Li.13a}, and anisotropic scattering~\cite{Pol.12a}, where scattering anisotropy was achieved by stretching the aerogel during growth producing a positive uniaxial strain.  In this Letter we report  our discovery  that  negative strain from  compression, contrary to the theory, stabilizes the isotropic phase and suppresses the A-phase. The competition between the strain and the magnetic field reveals a new critical point $(H_c, T_c)$ in the magnetic field-temperature plane.  Furthermore, we report that  the orientation of the angular momentum in this destabilized A-phase is perpendicular to the direction of strain  contrary to theoretical predictions~\cite{Vol.08,Sau.13}.  Consequently, the angular momentum has a continuous rotational symmetry with respect to the strain axis and  becomes a two-dimensional  superfluid glass.

\begin{figure}
\centerline{\includegraphics[height=0.40\textheight]{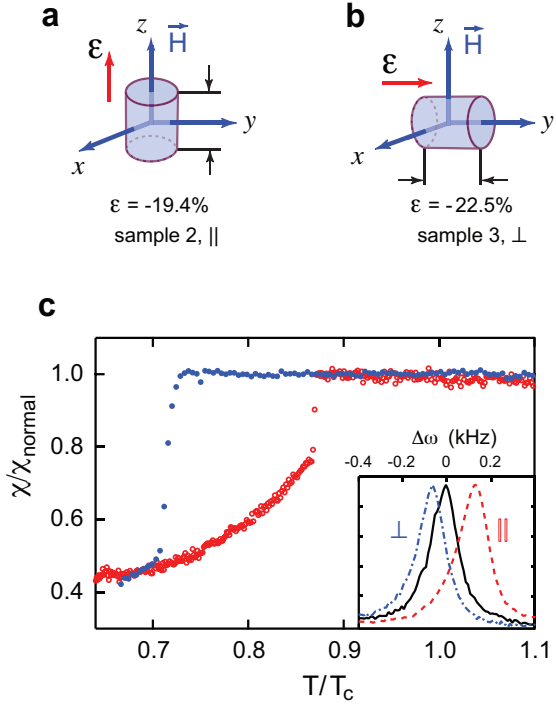}}
\caption{\label{fig1}(Color online).  a) and b)  Schematics of the experimental arrangements for samples 2 and 3.  Sample 2, $\bm{\varepsilon}$$\,=\,$$ -19.4$\%, has  strain axis parallel to the magnetic field.   For sample 3, $\bm{\varepsilon}$$\,=\,$$ -22.5$\%, the strain axis is perpendicular to the magnetic field. c) Liquid susceptibility normalized to the susceptibility of the normal state on warming (open red circles) and cooling (solid blue circles) versus reduced temperature for sample 3 in $H = 196$ mT. Similar behavior was observed in sample 2, but was omitted for clarity.  Inset: NMR spectra for $^3$He in aerogel: sample 3 in the normal state (black solid curve) and A-phase (blue dashed-dot curve); sample 2 in the A-phase (red dashed curve). The normal state spectra have the same line shape.  The spectra for samples 2 (3) were obtained in $H =  95.6 \, (79.2)$ mT respectively and have been scaled to a common field and temperature for  comparison.}
\end{figure}

In the present work we compare three different samples of nominally 98\% porosity silica aerogel in the shape of cylinders $4.0$ mm in diameter   with unstrained length of $5.1$ mm.  After their growth all samples were found to be isotropic and  homogeneous  using optical birefringence following established procedures~\cite{Pol.08} \footnote{A positively strained aerogel exhibits optical birefringence but becomes non-birefringent under compression at exactly the value expected for  compensation of the original growth-induced positive strain. The compression process is reversible, meaning that after the compression of an isotropic aerogel is released it reverts to being uniformly isotropic with its original dimensions.}.  Sample 1 was the subject of several previous reports~\cite{Pol.11,Li.12,Li.13a} where we identified the $p$-wave superfluid states to be  A and B-phases, with a suppressed order parameter amplitude, in a range of magnetic fields and pressures concluding that  for an isotropic unstrained aerogel in zero field there is but one superfluid state, irrespective of pressure and temperature, and this is the isotropic B-phase.  The superfluid transitions were sharp ~$\Delta T_c/T_c\lesssim0.2 \%$ and precisely reproduced in several cool downs.  This exact sample was then compressed to a strain $\bm{\varepsilon}$$\,=\,$$ -19.4$\%, directed along the field, $\bm{\varepsilon}$$\parallel$$\bm{H}$, Fig.~1a. We refer to this sample as sample 2.  Sample 3, was grown and characterized in the same way as was sample 1, and was compressed to a strain $\bm{\varepsilon}$$\,=\,$$ -22.5$\% oriented with $\bm{\varepsilon}$$\perp$$\bm{H}$, Fig.~1b.

Pulsed NMR measurements were performed at a pressure $P$\,=\,26.3 bar in magnetic fields ranging from $H$\,= \,49.1 to 196 mT. After an RF pulse that  tips the nuclear magnetization of the $^3$He atoms by an angle $\beta$ away from the external field,  Fourier transformation of the free induction decay signal was phase corrected to obtain  absorption spectra, as shown in the inset to Fig.~1c. 
The magnetic susceptibility, $\chi$, was determined from the numerical integral of the spectrum. The thermodynamic transition between superfluid states was precisely determined from the susceptibility discontinuity on warming (see Fig.~1c) displayed in phase diagrams in Fig.~2.   On cooling the supercooled transitions appear $\approx300\, \mu$K below $T_c$ for all samples and magnetic fields, as indicated in Fig.~1c.  The frequency shift of the spectrum, $\Delta\omega$, was calculated from the  first moment of each spectrum relative to the Larmor frequency,  $\omega_L$, given by the resonance frequency in the normal state.  From the spectra in the inset of Fig.~1c,  it is clear that $\bm{\varepsilon}$\,$\parallel$\,$\bm{H}$ has a positive shift and $\bm{\varepsilon}$\,$\perp$\,$\bm{H}$ has a negative shift in the A-phase.  We infer  that the orientation of the strain axis relative to the field controls the sign of the frequency shift; this implies specific directions for the angular momentum which we will discuss later in the context of theory~\cite{Vol.06, Sau.13}. 


\begin{figure}
\centerline{\includegraphics[height=0.26\textheight]{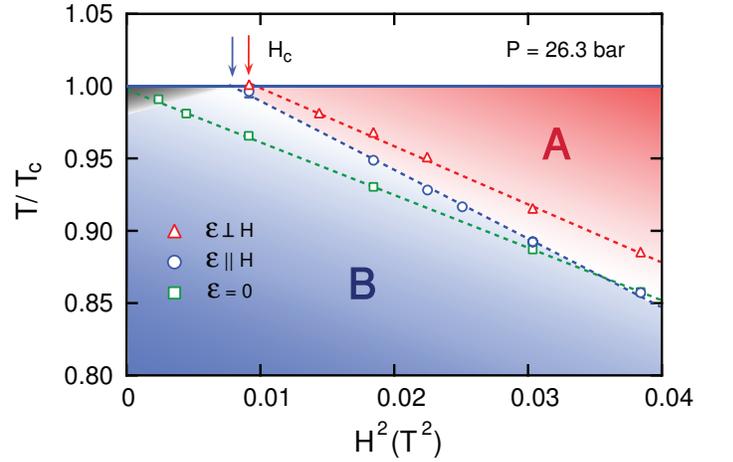}}
\caption{\label{fig2} (Color online).  Superfluid phase diagram $T_{BA}/T_c$ versus $H^2$ for  three samples at a pressure of $P = 26.3$ bar from warming experiments. Two of the three have negative strain, $\bm{\varepsilon}<0$, and one is isotropic, $\bm{\varepsilon}=0$.  All three have the common feature of a quadratic dependence on magnetic field for the B to A-phase transitions.  Remarkably, for negative strain there appears to be a critical field, $H_c$.}
\end{figure}

In sufficiently large magnetic fields, an equal-spin-pairing (ESP) state, {\it i.e.} having the same susceptibility as the normal state, is stabilized near $T_c$, Fig.~2.  For sample 1 we have established earlier~\cite{Pol.11} that this ESP phase is the A-phase. The equilibrium phase boundaries $T_{BA}(H)$ at $P=26.3$ bar from B to A-phases on warming,  are shown in Fig.~2  for all three samples as a function of magnetic field.  For the isotropic aerogel, $\bm{\varepsilon}$\,=\,0 (sample 1, green squares), the extrapolation of $T_{BA}$ for  $H^2 \rightarrow 0$ occurs precisely at $T_{BA}/T_c= 1$.   This is expected for a quadratic field contribution to the  Ginzburg-Landau free energy following  exactly the same  behavior as is well-established for  pure $^3$He  below the polycritical point,  $P<21$ bar~\cite{Ger.02,Hal.08}.  However, for the anisotropic aerogels, $\bm{\varepsilon}\,<\,0$ (sample 2, blue circles and sample 3, red triangles) the B-phase stability is enhanced with respect to the A-phase as a result of a positive offset of this quadratic field dependence described by,  
\begin{equation}
1-\frac{T_{BA}}{T_c}=g_{BA}\left( \frac{H^2-H^2_c}{H^2_0}\right) +\mathit{O}\left( \frac{H}{H_{0}}\right)^4,\label{1}\\
\end{equation}
\noindent
thereby defining a critical field  $H_c$ at $T_{BA}/T_c= 1$.  Here $H_c$ is 88.6 mT and $97.8$ mT for samples 2 and 3 respectively.   $H_0$ and $g_{BA}$ are superfluid constants from the GL theory \cite{Thu.98,Cho.07}.  

We have calculated the quasiparticle mean free path, $\lambda$,  by fitting  our results to Eq.~1 using the homogeneous isotropic scattering model~\cite{Thu.98,Sau.03}.  For sample 1, the isotropic aerogel,   $H_c$\,=\,0 and we find $\lambda = 210$ nm.  Similarly,  for the compressed aerogel samples 2 and 3 we have calculated $\lambda$ from the slope of the phase line in Fig.~2 to be $250$ and $220$ nm respectively, all of which are very reasonable values for a 98\% porous aerogel~\cite{Hal.08}.


\begin{figure*}
\centerline{\includegraphics[height=0.38\textheight]{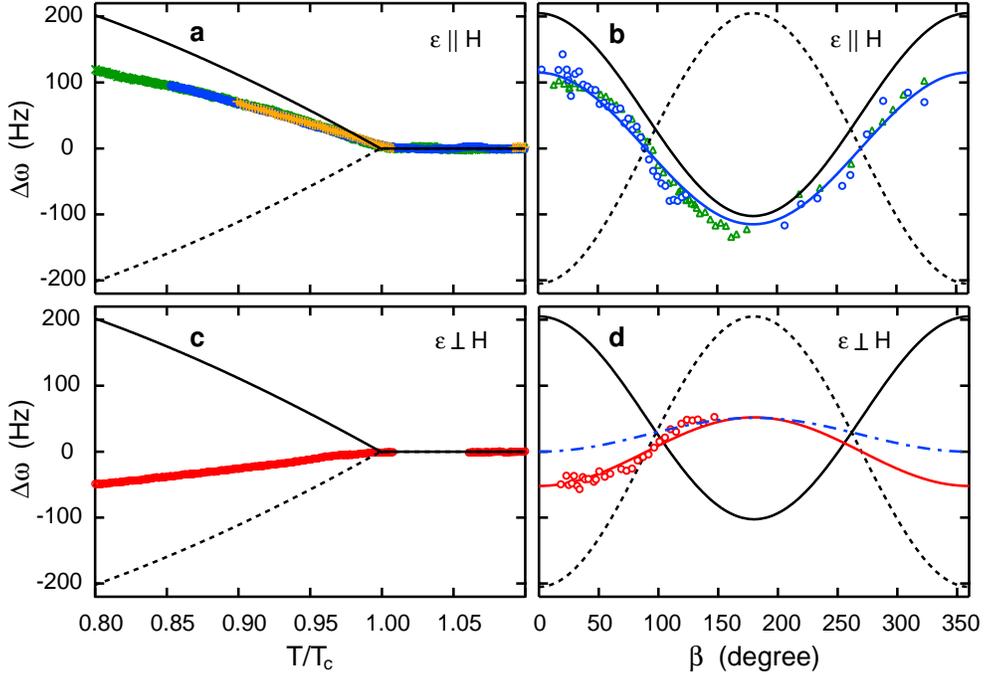}}
\caption{\label{fig3} (Color online). NMR frequency shifts as a function of temperature and tip angle. The magnitude of the shifts presented here has been scaled to a common field of $H = 196$ mT, and the tip angle dependence has been scaled to a common temperature of $T/T_c = 0.8$ based on our results from sample 1.  In all the panels, the dipole-locked configuration, $\el$\,$\perp$\,$\bm{H}$ is shown by the black solid curves, and the black dashed curves are  the  dipole-unlocked configuration, $\el$\,$\parallel$\,$\bm{H}$, with $\Omega_A$ taken from the measurements for sample 1~\cite{Pol.11}.  Comparison of frequency shifts in the A-phase: a, b) for $\bm{\varepsilon}$\,$\parallel$\,$\bm{H}$  and  c,d) $\bm{\varepsilon}$\,$\perp$\,$\bm{H}$.  a, c) Dependence on reduced temperature with small tip angle, $\beta < 20^{\circ}$. a) warming in $H=174$ mT, yellow crosses; cooling in $H=174$ mT, green x; cooling in $H=95.6$ mT, blue circles.  c) cooling in $H=79.2$ mT, red circles.  b, d) Dependence on tip angle after cooling from the normal state. b) $T/T_c=0.87$, $H=49.1$ mT, green squares; $T/T_c=0.82$, $H=196$ mT, blue circles. d) $T/T_c=0.78$, $H=95.6$ mT, red circles.  The blue and red solid curves are calculated for the easy-plane, 2D glass model, see text.}
\end{figure*}

The existence of a critical field, $H_c$, is unexpected.  It indicates that anisotropic impurity scattering with $\bm{\varepsilon}\,<\,0$ introduces a new additive term in the free energy which depends on strain.  Furthermore, contrary to theory~\cite{Thu.98,Aoy.06,Fom.03}, it appears that aerogel anisotropy induced by uniaxial compression enhances the B-phase stability relative to the A-phase,  countering the effect of an applied magnetic field.  

Based on symmetry arguments there are two possibilities for the direction of the angular momentum, $\el$, in the presence of a large uniaxial strain which determine the sign of the frequency shift in the anisotropic A-phase.  One model requires that the angular momentum be parallel to the strain $\el$\,$\parallel$\,$\bm{\varepsilon}$, called the easy-axis model, which was predicted for $ \bm{\varepsilon} < 0 $ by two different theories~\cite{Vol.06, Sau.13}.  The other, the easy-plane model, allows the angular momentum to be in a plane perpendicular to the strain, $\el$\,$\perp$\,$\bm{\varepsilon}$, not favored by either theory if $ \bm{\varepsilon} $\,$< $\,$0 $.

Using  NMR frequency shift measurements as a function of temperature and NMR tip angle we have investigated the orientation of the angular momentum for these two models. The dipole energy in the A-phase is~\cite{Vol.90}, $F_D=-\frac{1}{2}\Omega^2_A\frac{\chi_A}{\gamma^2}\left ( \el \cdot \bm{\hat{d}}\right )^2$ where $\Omega_A$ is the longitudinal resonance frequency, $\gamma$ is the gyromagnetic ratio, $\chi_A$ is the nuclear magnetic susceptibility and $\bm{\hat{d}}$ is a spin-space vector constrained to be perpendicular to the spin angular momentum, $\bm{\hat{s}}$, while minimizing $F_D$. The relative orientation of the A-phase order parameters, $\el$ and $\bm{\hat{d}}$, can be parametrized by two angles $\theta$ and $\phi$, where $\theta$ is the angle between $\el$ and the magnetic field $\bm{H}$; and $\phi$ is the angle between $\bm{\hat{d}}$ and the projection of $\el$ in the plane perpendicular to the field. We note that $\Omega_A$ should be approximately the same for all three samples since the quasiparticle mean free paths are very similar ~\cite{Hal.08}. The dependence of the frequency shift on tip angle, $\beta$, is given by \cite{Yur.93b,Vol.06}:

       \begin{eqnarray}
      \Delta\omega&=&\frac{{\Omega_A}^2}{2\omega_L} \left(-\cos\beta + \left(\frac{7}{4}\cos\beta+\frac{1}{4}\right) \left \langle \sin^2 \theta(\bm{r}) \right \rangle \right) \hspace{20pt}\notag\\
      &-&\frac{{\Omega_A}^2}{2\omega_L}\left(\frac{1}{2} \left(1+\cos\beta\right) \left \langle\sin^2\phi(\bm{r})\right \rangle \left \langle \sin^2 \theta(\bm{r}) \right \rangle\right), \hspace{20pt}\label{2}      
      \end{eqnarray} 
\noindent
where $\left \langle \sin^2 \theta(\bm{r}) \right \rangle$ and $\left \langle \sin^2 \phi(\bm{r}) \right \rangle$ are  spatial averages over a dipole length $\xi_D$ at location $\bm{r}$.  According to the easy-axis model, $\bm{\varepsilon}\parallel \el$, we have $\theta = 0$ for sample 2 ($\bm{\varepsilon}$\,$\parallel$\,$\bm{H}$) and $\theta = \frac{\pi}{2}$, $\phi = 0$ for sample 3 ($\bm{\varepsilon}$\,$\perp$\,$\bm{H}$). For these two orientations of the strain relative to the magnetic field Eq.~2 becomes:

\begin{eqnarray}
\Delta\omega_{\bm{\varepsilon}\parallel\bm{H}}\left(\beta\right) &=& -\frac{{\Omega_A}^2}{2\omega_L}\cos\beta, \label{3}\\
\Delta\omega_{\bm{\varepsilon}\perp\bm{H}} \left(\beta\right) &=& \frac{{\Omega_A}^2}{2\omega_L}\frac{\left(3\cos\beta+1\right)}{4}. \label{4}						
\end{eqnarray}

\noindent

Our frequency shift data in the small tip angle limit, Fig. 3a and 3c have  signs opposite to Eqs.~3 and 4.  Furthermore, the magnitude of the shifts observed are smaller than the predictions for the easy-axis model, which are the maximum (minimum) possible shifts for the A-phase often referred to as dipole-locked (unlocked) configurations~\cite{Vol.90}, shown as black solid (dashed) curves. Consequently the easy-axis model  is incorrect. 

For the easy-plane model,  $\el$\,$\perp$\,$\bm{\varepsilon}$, there is a continuous symmetry for the angular momentum in the plane perpendicular to the strain.  This will lead to a 2-dimensional (2D) superfluid glass phase similar to the 3D glass state observed for sample 1~\cite{Li.13a}.  In this case,  for sample 2, $\theta = \frac{\pi}{2}$ and $\left\langle\sin^2\theta\right\rangle = 1$; whereas for sample 3, $0 < \theta < \frac{\pi}{2}$ and $\left\langle\sin^2\theta\right\rangle = \frac{1}{2}$.  To calculate $\Delta \omega$ we must constrain $\phi$ and we consider two possible scenarios.  In the first scenario,  the $\bm{\hat{d}}$-vector is also disordered on a length scale smaller than a dipole length then, $0 < \phi < 2\pi$ and $\left\langle\sin^2\phi\right\rangle = \frac{1}{2}$, and Eq.~2 gives,

\begin{eqnarray}
\Delta\omega_{\bm{\varepsilon}\parallel\bm{H}}\left(\beta\right) &=& \frac{{\Omega_A}^2}{2\omega_L} \left(\frac{1}{2}\cos\beta\right), \\
\Delta\omega_{\bm{\varepsilon}\perp\bm{H}}\left(\beta\right) &=& \frac{{\Omega_A}^2}{2\omega_L}\left(-\frac{1}{4}\cos\beta\right).						
\end{eqnarray}

Our calculations for this case, shown as solid blue and red curves in Fig.~3b and 3d, are  in excellent agreement with our measurements as a function of tip angle and coincide with the measured temperature dependences for both orientations of the strain relative to the magnetic field, Fig. 3a and 3c. From this calculation, we obtained the value of $\Omega_A$ at $T/T_c=0.8$, $\Omega_A = 51.0$, $54.0$ and $51.2$ kHz for sample 1, 2 and 3 respectively, within the combined measurement errors of $10\%$. The consistency in the values of $\Omega_A$ strongly supports this scenario. 

For the second scenario, the $\bm{\hat{d}}$-vector has a uniform orientation on length scales longer than the dipole length.  Then Eq.~5 holds for sample 2; however, for sample 3, $\phi = 0$ and $\left\langle\sin^2\phi\right\rangle = 0$, and from Eq.~2,

\begin{eqnarray}
\Delta\omega_{\bm{\varepsilon}\perp\bm{H}}\left(\beta\right) &=& \frac{{\Omega_A}^2}{8\omega_L}\left(1-\cos\beta\right).						
\end{eqnarray}

\noindent
 At small tip angles the  frequency shift from Eq.~7 is zero for all temperatures, inconsistent with the data in Fig.~3c. Additionally, the tip angle dependence  from  Eq.~7, shown as a blue dash-dotted curve in Fig.~3d, is  inconsistent with the data indicating that the first scenario better describes the $\bm{\hat{d}}$-vector orientation. We infer that the easy-plane model correctly describes  the orientation of the angular momentum to be constrained to a plane perpendicular to the strain-axis for a compressed  aerogel, giving rise to a 2D glass phase in that plane. 

Recent measurements of the superfluid density,  $\rho_s/\rho$, in zero applied field with a  torsional oscillator using a similarly prepared aerogel sample with $10\%$ compression ~\cite{Ben.11} found evidence of a stable phase just below $T_c$. In the magnetic field-temperature plane, such a stable phase might occupy the shaded area, Fig.~2, resulting in a positive slope of the phase boundary to the B-phase. We cannot make a direct comparison with  these results since our NMR experiments have not been performed in sufficiently low magnetic fields. However, according to the Clausius-Clapeyron relation such a stable phase could not be an ESP state.

In summary, we have investigated the nature of superfluid $^3$He-A in two uniformly anisotropic aerogel samples with negative strain achieved with uniaxial compression of $\sim$20\% and we have compared them with isotropic aerogel. We discovered a critical field corresponding to a new critical point ($T_c,H_c$) at a pressure of $P=26.3$ bar and that negative strain enhances the stability of the isotropic B-phase over the A-phase in a magnetic field, in contrast with existing theories ~\cite{Thu.98,Aoy.06,Fom.03}.  Furthermore, the angular momentum in the A-phase is not aligned with the strain axis as has been predicted~\cite{Vol.08,Sau.13}, rather it forms a 2D glass phase in the plane perpendicular to the strain.

\begin{acknowledgments}
We are grateful to V.V. Dmitriev, J.M. Parpia, J.A Sauls, G.E. Volovik for helpful discussion and for support from the National Science Foundation, DMR-1103625.
\end{acknowledgments}


\begin{thebibliography}{19}%
\makeatletter
\providecommand \@ifxundefined [1]{%
 \@ifx{#1\undefined}
}%
\providecommand \@ifnum [1]{%
 \ifnum #1\expandafter \@firstoftwo
 \else \expandafter \@secondoftwo
 \fi
}%
\providecommand \@ifx [1]{%
 \ifx #1\expandafter \@firstoftwo
 \else \expandafter \@secondoftwo
 \fi
}%
\providecommand \natexlab [1]{#1}%
\providecommand \enquote  [1]{``#1''}%
\providecommand \bibnamefont  [1]{#1}%
\providecommand \bibfnamefont [1]{#1}%
\providecommand \citenamefont [1]{#1}%
\providecommand \href@noop [0]{\@secondoftwo}%
\providecommand \href [0]{\begingroup \@sanitize@url \@href}%
\providecommand \@href[1]{\@@startlink{#1}\@@href}%
\providecommand \@@href[1]{\endgroup#1\@@endlink}%
\providecommand \@sanitize@url [0]{\catcode `\\12\catcode `\$12\catcode
  `\&12\catcode `\#12\catcode `\^12\catcode `\_12\catcode `\%12\relax}%
\providecommand \@@startlink[1]{}%
\providecommand \@@endlink[0]{}%
\providecommand \url  [0]{\begingroup\@sanitize@url \@url }%
\providecommand \@url [1]{\endgroup\@href {#1}{\urlprefix }}%
\providecommand \urlprefix  [0]{URL }%
\providecommand \Eprint [0]{\href }%
\providecommand \doibase [0]{http://dx.doi.org/}%
\providecommand \selectlanguage [0]{\@gobble}%
\providecommand \bibinfo  [0]{\@secondoftwo}%
\providecommand \bibfield  [0]{\@secondoftwo}%
\providecommand \translation [1]{[#1]}%
\providecommand \BibitemOpen [0]{}%
\providecommand \bibitemStop [0]{}%
\providecommand \bibitemNoStop [0]{.\EOS\space}%
\providecommand \EOS [0]{\spacefactor3000\relax}%
\providecommand \BibitemShut  [1]{\csname bibitem#1\endcsname}%
\let\auto@bib@innerbib\@empty
\bibitem [{\citenamefont {Pollanen}\ \emph {et~al.}(2012)\citenamefont
  {Pollanen}, \citenamefont {Li}, \citenamefont {Collett}, \citenamefont
  {Gannon}, \citenamefont {Halperin},\ and\ \citenamefont {Sauls}}]{Pol.12a}%
  \BibitemOpen
  \bibfield  {author} {\bibinfo {author} {\bibfnamefont {J.}~\bibnamefont
  {Pollanen}}, \bibinfo {author} {\bibfnamefont {J.~I.~A.}\ \bibnamefont {Li}},
  \bibinfo {author} {\bibfnamefont {C.~A.}\ \bibnamefont {Collett}}, \bibinfo
  {author} {\bibfnamefont {W.~J.}\ \bibnamefont {Gannon}}, \bibinfo {author}
  {\bibfnamefont {W.~P.}\ \bibnamefont {Halperin}}, \ and\ \bibinfo {author}
  {\bibfnamefont {J.~A.}\ \bibnamefont {Sauls}},\ }\href@noop {} {\bibfield
  {journal} {\bibinfo  {journal} {Nature Phys.}\ }\textbf {\bibinfo {volume}
  {8}},\ \bibinfo {pages} {317} (\bibinfo {year} {2012})}\BibitemShut {NoStop}%
\bibitem [{\citenamefont {Pollanen}\ \emph {et~al.}(2011)\citenamefont
  {Pollanen}, \citenamefont {Li}, \citenamefont {Collett}, \citenamefont
  {Gannon},\ and\ \citenamefont {Halperin}}]{Pol.11}%
  \BibitemOpen
  \bibfield  {author} {\bibinfo {author} {\bibfnamefont {J.}~\bibnamefont
  {Pollanen}}, \bibinfo {author} {\bibfnamefont {J.~I.~A.}\ \bibnamefont {Li}},
  \bibinfo {author} {\bibfnamefont {C.~A.}\ \bibnamefont {Collett}}, \bibinfo
  {author} {\bibfnamefont {W.~J.}\ \bibnamefont {Gannon}}, \ and\ \bibinfo
  {author} {\bibfnamefont {W.~P.}\ \bibnamefont {Halperin}},\ }\href@noop {}
  {\bibfield  {journal} {\bibinfo  {journal} {Phys. Rev. Lett.}\ }\textbf
  {\bibinfo {volume} {107}},\ \bibinfo {pages} {195301} (\bibinfo {year}
  {2011})}\BibitemShut {NoStop}%
\bibitem [{\citenamefont {Thuneberg}\ \emph {et~al.}(1998)\citenamefont
  {Thuneberg}, \citenamefont {Yip}, \citenamefont {Fogelstr{\"o}m},\ and\
  \citenamefont {Sauls}}]{Thu.98}%
  \BibitemOpen
  \bibfield  {author} {\bibinfo {author} {\bibfnamefont {E.~V.}\ \bibnamefont
  {Thuneberg}}, \bibinfo {author} {\bibfnamefont {S.~K.}\ \bibnamefont {Yip}},
  \bibinfo {author} {\bibfnamefont {M.}~\bibnamefont {Fogelstr{\"o}m}}, \ and\
  \bibinfo {author} {\bibfnamefont {J.~A.}\ \bibnamefont {Sauls}},\ }\href@noop
  {} {\bibfield  {journal} {\bibinfo  {journal} {Phys. Rev. Lett.}\ }\textbf
  {\bibinfo {volume} {80}},\ \bibinfo {pages} {2861} (\bibinfo {year}
  {1998})}\BibitemShut {NoStop}%
\bibitem [{\citenamefont {Gervais}\ \emph {et~al.}(2002)\citenamefont
  {Gervais}, \citenamefont {Yawata}, \citenamefont {Mulders},\ and\
  \citenamefont {Halperin}}]{Ger.02}%
  \BibitemOpen
  \bibfield  {author} {\bibinfo {author} {\bibfnamefont {G.}~\bibnamefont
  {Gervais}}, \bibinfo {author} {\bibfnamefont {K.}~\bibnamefont {Yawata}},
  \bibinfo {author} {\bibfnamefont {N.}~\bibnamefont {Mulders}}, \ and\
  \bibinfo {author} {\bibfnamefont {W.~P.}\ \bibnamefont {Halperin}},\
  }\href@noop {} {\bibfield  {journal} {\bibinfo  {journal} {Phys. Rev. B.}\
  }\textbf {\bibinfo {volume} {66}},\ \bibinfo {pages} {054528} (\bibinfo
  {year} {2002})}\BibitemShut {NoStop}%
\bibitem [{\citenamefont {Li}\ \emph {et~al.}(2013)\citenamefont {Li},
  \citenamefont {Pollanen}, \citenamefont {Zimmerman}, \citenamefont {Collett},
  \citenamefont {Gannon},\ and\ \citenamefont {Halperin}}]{Li.13a}%
  \BibitemOpen
  \bibfield  {author} {\bibinfo {author} {\bibfnamefont {J.~I.~A.}\
  \bibnamefont {Li}}, \bibinfo {author} {\bibfnamefont {J.}~\bibnamefont
  {Pollanen}}, \bibinfo {author} {\bibfnamefont {A.~M.}\ \bibnamefont
  {Zimmerman}}, \bibinfo {author} {\bibfnamefont {C.~A.}\ \bibnamefont
  {Collett}}, \bibinfo {author} {\bibfnamefont {W.~J.}\ \bibnamefont {Gannon}},
  \ and\ \bibinfo {author} {\bibfnamefont {W.~P.}\ \bibnamefont {Halperin}},\
  }\href@noop {} {\bibfield  {journal} {\bibinfo  {journal} {Nature Physics}\
  }\textbf {\bibinfo {volume} {9}},\ \bibinfo {pages} {775} (\bibinfo {year}
  {2013})}\BibitemShut {NoStop}%
\bibitem [{\citenamefont {Volovik}(2008)}]{Vol.08}%
  \BibitemOpen
  \bibfield  {author} {\bibinfo {author} {\bibfnamefont {G.~E.}\ \bibnamefont
  {Volovik}},\ }\href@noop {} {\bibfield  {journal} {\bibinfo  {journal} {J.
  Low Temp. Phys.}\ }\textbf {\bibinfo {volume} {150}},\ \bibinfo {pages} {453}
  (\bibinfo {year} {2008})}\BibitemShut {NoStop}%
\bibitem [{\citenamefont {Sauls}(2013)}]{Sau.13}%
  \BibitemOpen
  \bibfield  {author} {\bibinfo {author} {\bibfnamefont {J.~A.}\ \bibnamefont
  {Sauls}},\ }\href@noop {} {\bibfield  {journal} {\bibinfo  {journal} {Phys.
  Rev. B.}\ }\textbf {\bibinfo {volume} {88}},\ \bibinfo {pages} {214503}
  (\bibinfo {year} {2013})}\BibitemShut {NoStop}%
\bibitem [{\citenamefont {Pollanen}\ \emph {et~al.}(2008)\citenamefont
  {Pollanen}, \citenamefont {Shirer}, \citenamefont {Blinstein}, \citenamefont
  {Davis}, \citenamefont {Choi}, \citenamefont {Lippman}, \citenamefont
  {Lurio},\ and\ \citenamefont {Halperin}}]{Pol.08}%
  \BibitemOpen
  \bibfield  {author} {\bibinfo {author} {\bibfnamefont {J.}~\bibnamefont
  {Pollanen}}, \bibinfo {author} {\bibfnamefont {K.~R.}\ \bibnamefont
  {Shirer}}, \bibinfo {author} {\bibfnamefont {S.}~\bibnamefont {Blinstein}},
  \bibinfo {author} {\bibfnamefont {J.~P.}\ \bibnamefont {Davis}}, \bibinfo
  {author} {\bibfnamefont {H.}~\bibnamefont {Choi}}, \bibinfo {author}
  {\bibfnamefont {T.~M.}\ \bibnamefont {Lippman}}, \bibinfo {author}
  {\bibfnamefont {L.~B.}\ \bibnamefont {Lurio}}, \ and\ \bibinfo {author}
  {\bibfnamefont {W.~P.}\ \bibnamefont {Halperin}},\ }\href@noop {} {\bibfield
  {journal} {\bibinfo  {journal} {J. Non-Crystalline Solids}\ }\textbf
  {\bibinfo {volume} {354}},\ \bibinfo {pages} {4668} (\bibinfo {year}
  {2008})}\BibitemShut {NoStop}%
\bibitem [{Note1()}]{Note1}%
  \BibitemOpen
  \bibinfo {note} {A positively strained aerogel exhibits optical birefringence
  but becomes non-birefringent under compression at exactly the value expected
  for compensation of the original growth-induced positive strain. The
  compression process is reversible, meaning that after the compression of an
  isotropic aerogel is released it reverts to being uniformly isotropic with
  its original dimensions.}\BibitemShut {Stop}%
\bibitem [{\citenamefont {Li}\ \emph {et~al.}(2012)\citenamefont {Li},
  \citenamefont {Pollanen}, \citenamefont {Collett}, \citenamefont {Gannon},\
  and\ \citenamefont {Halperin}}]{Li.12}%
  \BibitemOpen
  \bibfield  {author} {\bibinfo {author} {\bibfnamefont {J.~I.~A.}\
  \bibnamefont {Li}}, \bibinfo {author} {\bibfnamefont {J.}~\bibnamefont
  {Pollanen}}, \bibinfo {author} {\bibfnamefont {C.~A.}\ \bibnamefont
  {Collett}}, \bibinfo {author} {\bibfnamefont {W.~J.}\ \bibnamefont {Gannon}},
  \ and\ \bibinfo {author} {\bibfnamefont {W.~P.}\ \bibnamefont {Halperin}},\
  }\href {\doibase 10.1088/1742-6596/400/1/012039} {\bibfield  {journal}
  {\bibinfo  {journal} {J. Phys.: Conf. Ser.}\ }\textbf {\bibinfo {volume}
  {400}},\ \bibinfo {pages} {012039} (\bibinfo {year} {2012})}\BibitemShut
  {NoStop}%
\bibitem [{\citenamefont {Volovik}(2006)}]{Vol.06}%
  \BibitemOpen
  \bibfield  {author} {\bibinfo {author} {\bibfnamefont {G.~E.}\ \bibnamefont
  {Volovik}},\ }\href@noop {} {\bibfield  {journal} {\bibinfo  {journal} {JETP
  Lett.}\ }\textbf {\bibinfo {volume} {84}},\ \bibinfo {pages} {533} (\bibinfo
  {year} {2006})}\BibitemShut {NoStop}%
\bibitem [{\citenamefont {Halperin}\ \emph {et~al.}(2008)\citenamefont
  {Halperin}, \citenamefont {Choi}, \citenamefont {Davis},\ and\ \citenamefont
  {Pollanen}}]{Hal.08}%
  \BibitemOpen
  \bibfield  {author} {\bibinfo {author} {\bibfnamefont {W.~P.}\ \bibnamefont
  {Halperin}}, \bibinfo {author} {\bibfnamefont {H.}~\bibnamefont {Choi}},
  \bibinfo {author} {\bibfnamefont {J.~P.}\ \bibnamefont {Davis}}, \ and\
  \bibinfo {author} {\bibfnamefont {J.}~\bibnamefont {Pollanen}},\ }\href@noop
  {} {\bibfield  {journal} {\bibinfo  {journal} {J. Phys. Soc. Jpn.}\ }\textbf
  {\bibinfo {volume} {77}},\ \bibinfo {pages} {111002} (\bibinfo {year}
  {2008})}\BibitemShut {NoStop}%
\bibitem [{\citenamefont {Choi}\ \emph {et~al.}(2007)\citenamefont {Choi},
  \citenamefont {Davis}, \citenamefont {Pollanen}, \citenamefont {Haard},\ and\
  \citenamefont {Halperin}}]{Cho.07}%
  \BibitemOpen
  \bibfield  {author} {\bibinfo {author} {\bibfnamefont {H.}~\bibnamefont
  {Choi}}, \bibinfo {author} {\bibfnamefont {J.~P.}\ \bibnamefont {Davis}},
  \bibinfo {author} {\bibfnamefont {J.}~\bibnamefont {Pollanen}}, \bibinfo
  {author} {\bibfnamefont {T.~M.}\ \bibnamefont {Haard}}, \ and\ \bibinfo
  {author} {\bibfnamefont {W.~P.}\ \bibnamefont {Halperin}},\ }\href@noop {}
  {\bibfield  {journal} {\bibinfo  {journal} {Phys. Rev. B}\ }\textbf {\bibinfo
  {volume} {75}},\ \bibinfo {pages} {174503} (\bibinfo {year}
  {2007})}\BibitemShut {NoStop}%
\bibitem [{\citenamefont {Sauls}\ and\ \citenamefont {Sharma}(2003)}]{Sau.03}%
  \BibitemOpen
  \bibfield  {author} {\bibinfo {author} {\bibfnamefont {J.~A.}\ \bibnamefont
  {Sauls}}\ and\ \bibinfo {author} {\bibfnamefont {P.}~\bibnamefont {Sharma}},\
  }\href@noop {} {\bibfield  {journal} {\bibinfo  {journal} {Phys. Rev. B}\
  }\textbf {\bibinfo {volume} {68}},\ \bibinfo {pages} {224502} (\bibinfo
  {year} {2003})}\BibitemShut {NoStop}%
\bibitem [{\citenamefont {Aoyama}\ and\ \citenamefont {Ikeda}(2006)}]{Aoy.06}%
  \BibitemOpen
  \bibfield  {author} {\bibinfo {author} {\bibfnamefont {K.}~\bibnamefont
  {Aoyama}}\ and\ \bibinfo {author} {\bibfnamefont {R.}~\bibnamefont {Ikeda}},\
  }\href@noop {} {\bibfield  {journal} {\bibinfo  {journal} {Phys. Rev. B}\
  }\textbf {\bibinfo {volume} {73}},\ \bibinfo {pages} {060504(R)} (\bibinfo
  {year} {2006})}\BibitemShut {NoStop}%
\bibitem [{\citenamefont {Fomin}(2003)}]{Fom.03}%
  \BibitemOpen
  \bibfield  {author} {\bibinfo {author} {\bibfnamefont {I.~A.}\ \bibnamefont
  {Fomin}},\ }\href@noop {} {\bibfield  {journal} {\bibinfo  {journal} {JETP
  Lett.}\ }\textbf {\bibinfo {volume} {77}},\ \bibinfo {pages} {240} (\bibinfo
  {year} {2003})}\BibitemShut {NoStop}%
\bibitem [{\citenamefont {Vollhardt}\ and\ \citenamefont
  {W{\"o}lfle}(1990)}]{Vol.90}%
  \BibitemOpen
  \bibfield  {author} {\bibinfo {author} {\bibfnamefont {D.}~\bibnamefont
  {Vollhardt}}\ and\ \bibinfo {author} {\bibfnamefont {P.}~\bibnamefont
  {W{\"o}lfle}},\ }\href@noop {} {\emph {\bibinfo {title} {The Superfluid
  Phases of Helium 3}}}\ (\bibinfo  {publisher} {Taylor and Francis},\ \bibinfo
  {year} {1990})\BibitemShut {NoStop}%
\bibitem [{\citenamefont {Bunkov}\ and\ \citenamefont
  {Volovik}(1993)}]{Yur.93b}%
  \BibitemOpen
  \bibfield  {author} {\bibinfo {author} {\bibfnamefont {Y.~M.}\ \bibnamefont
  {Bunkov}}\ and\ \bibinfo {author} {\bibfnamefont {G.~E.}\ \bibnamefont
  {Volovik}},\ }\href@noop {} {\bibfield  {journal} {\bibinfo  {journal}
  {Europhys. Lett.}\ }\textbf {\bibinfo {volume} {21}},\ \bibinfo {pages} {837}
  (\bibinfo {year} {1993})}\BibitemShut {NoStop}%
\bibitem [{\citenamefont {Bennett}\ \emph {et~al.}(2011)\citenamefont
  {Bennett}, \citenamefont {Zhelev}, \citenamefont {Smith}, \citenamefont
  {Pollanen}, \citenamefont {Halperin},\ and\ \citenamefont {Parpia}}]{Ben.11}%
  \BibitemOpen
  \bibfield  {author} {\bibinfo {author} {\bibfnamefont {R.~G.}\ \bibnamefont
  {Bennett}}, \bibinfo {author} {\bibfnamefont {N.}~\bibnamefont {Zhelev}},
  \bibinfo {author} {\bibfnamefont {E.~N.}\ \bibnamefont {Smith}}, \bibinfo
  {author} {\bibfnamefont {J.}~\bibnamefont {Pollanen}}, \bibinfo {author}
  {\bibfnamefont {W.~P.}\ \bibnamefont {Halperin}}, \ and\ \bibinfo {author}
  {\bibfnamefont {J.~M.}\ \bibnamefont {Parpia}},\ }\href@noop {} {\bibfield
  {journal} {\bibinfo  {journal} {Phys. Rev. Lett.}\ }\textbf {\bibinfo
  {volume} {107}},\ \bibinfo {pages} {235504} (\bibinfo {year}
  {2011})}\BibitemShut {NoStop}%
\end{thebibliography}
\end{document}